\documentclass[10pt]{iopart}
\usepackage{iopams,texdraw}  
\begin{document}

\title[Critical exponents for isotropic Lifshitz
points\ldots]{Critical, crossover, and correction-to-scaling exponents
  for isotropic Lifshitz points to order
  $\boldsymbol{(8-d)^2}$%
}

\author{H.~W. Diehl\dag\ and M. Shpot\ddag}

\address{\dag\ Fachbereich Physik, Universit{\"a}t
Essen, D-45117 Essen, Federal Republic of Germany} 

\address{\ddag\ Institute for Condensed Matter Physics, 79011 Lviv,
  Ukraine}

\begin{abstract}
  A two-loop renormalization group analysis of the critical behaviour
  at an isotropic Lifshitz point is presented. Using dimensional
  regularization and minimal subtraction of poles, we obtain the
  expansions of the critical exponents $\nu$ and $\eta$, the crossover
  exponent $\varphi$, as well as the (related) wave-vector exponent
  $\beta_q$, and the correction-to-scaling exponent $\omega$ to second order in
  $\epsilon_8=8-d$.  These are compared with the authors' recent
  $\epsilon$-expansion results [{\it Phys.\ Rev.\ B} {\bf 62} (2000)
  12338; {\it Nucl.\ Phys.\ B} {\bf 612} (2001) 340] for the general
  case of an $m$-axial Lifshitz point. It is shown that the expansions
  obtained here by a direct calculation for the isotropic ($m=d$)
  Lifshitz point all follow from the latter upon setting
  $m=8-\epsilon_8$.  This is so despite recent claims to the contrary
  by de Albuquerque and Leite [{\it J.\ Phys.\ A} {\bf 35} (2002) 1807].
\end{abstract}




\section{Introduction}\label{sec:Intro}

The concept of a Lifshitz point, introduced in 1975 \cite{HLS75b}, has
attracted considerable attention during the past 25 years%
\footnote[3]{For reviews and extensive lists of references, see
\cite{Hor80,Sel92,DS00a,SD01}.
The papers \cite{DS00a,SD01} contain lists of the most recent references
on the subject.}.
Recently there has been renewed interest in the critical behaviour at
such points. In particular, field-theory approaches have been utilized
to determine the dimensionality expansions of various critical
indices needed to characterize the critical behaviour at $m$-axial
Lifshitz points \cite{DS00a,SD01,MC98,MC99,DS01b}.

We assume that the
order-parameter symmetry which is spontaneously broken on the
low-temperature side of the critical line on which the Lifshitz point
is located is $O(n)$, and that the wave-vector instability which sets in
at this point is isotropic in the $m$-dimensional subspace of
$\mathbb{R}^d$. A familiar model representing the universality class
of $d$-dimensional systems with short-range interactions and an
$n$-component order-parameter field $\bphi(\boldsymbol{x})$ at such an
$(m,d,n)$-Lifshitz point is described by the Hamiltonian
\begin{equation}\label{eq:Ham}
\fl
{\mathcal{H}}={\int}\!{d^d}x
{\left\{
{\sigma_0\over 2}\,{(\triangle_\parallel
 \bphi )}^2
+\frac{1}{2}\,{({{\nabla}_\perp}
 \bphi )}^2
+{\rho_0\over 2}\,{({{\nabla}_\parallel}
 \bphi )}^2
+{\tau_0\over 2}\,
\bphi^2+\frac{u_0}{4!}\,|\bphi |^4\right\}}\;.
\end{equation}
Here the position vector
$\boldsymbol{x}=(\boldsymbol{x}_\|,\boldsymbol{x}_\perp)$ has an
$m$-dimensional parallel component $\boldsymbol{x}_\|$ and a
$(d{-}m)$-dimensional perpendicular one $\boldsymbol{x}_\perp$. In the
Landau approximation, the Lifshitz point is located at
$\tau_0=\rho_0=0$; this approximation holds arbitrarily close to it
for values of $d$ exceeding the upper critical dimension, which is
$d^*(m)=4+m/2$ for $m\le 8$ \cite{HLS75b}.

Interest in the model (\ref{eq:Ham}) started already in 1975
\cite{HLS75b,Muk77,HB78,SG78}. Unfortunately, early analyses
\cite{HLS75b,Muk77,HB78,SG78} based on Wilson's momentum shell
recursion relations were rather incomplete and produced a
long-standing controversy: The series expansions to second order in
$\epsilon=d^*(m)-d$ one group of authors derived for the correlation
exponents $\eta_{l2}$ and $\eta_{l4}$ for general values
of $m$ \cite{Muk77} (or for $m=1$ \cite{HB78}) disagreed with those
obtained by another one \cite{SG78} for the special values $m=2$ and
$m=6$. This state was certainly very unsatisfactory because the class
of models defined via the Hamiltonian (\ref{eq:Ham}) is not only
interesting in its own right, but provides a natural, simple generalization
of the standard class of $n$-component $|\bphi|^4$-models (to which it
reduces when $m=0$).  Moreover, besides representing universality
classes, these models are of a prototypical nature in that they
exhibit \emph{anisotropic scale invariance}, offering excellent
possibilities for investigating the question whether and under what
conditions scale invariance might give rise to further invariances
analogous to Schr{\"o}dinger invariance or conformal symmetries
\cite{Hen97,PH01}. Clearly, their thorough theoretical understanding
is highly desirable.
  
In two recent papers \cite{DS00a,SD01}, we have been able to perform a
field-theoretic two-loop renormalization group (RG) analysis of the
model (\ref{eq:Ham}) near its upper critical dimension $d^*(m)$ for
general values of
$m$ with $0< m< 8$. The results yield the $\epsilon$ expansions to
second order of all critical, crossover, wave-vector, and
correction-to-scaling exponents about any point $(m,d^*(m))$ on the
critical line $d^*(m)$. The expansion coefficients involve four
well-defined single integrals $j_\phi(m)$, $j_\sigma(m)$, $j_\rho(m)$,
and $J_u(m)$, which we have been able to evaluate analytically for the
special values $m=0$, 2, 6, and 8, and for other values of $m$ by
numerical integration. An appealing feature of these results is that
\emph{they include both isotropic cases, namely that of the usual
critical point ($m=0$) and that of the isotropic Lifshitz point
($m=d=8-\epsilon_8$)}.

The requirement that our general $m$-dependent expressions reduce to
the correct results for these simpler cases in which only the
perpendicular or parallel parts of space remain provides crucial
checks. These issues were briefly discussed already in our previous
work \cite{SD01}, where also a direct two-loop RG calculation for the
isotropic case $m=d$ was announced but not described. Our aim here is
to fill this gap by presenting details of such an analysis and a
thorough discussion of the consistency of its results with our
previous findings for general $m$.%
\footnote{Such a calculation is analogous to the one for the
  standard $|\bphi|^4$ theory ($m=0$), which can be found in many
  textbooks like \cite{Ami84}. We had originally felt that the
  publication of this calculation was unnecessary. The fact that
  reference \cite{dAL02} got published and the reaction of many
  colleagues, who suggested that we write a clarifying paper, made us
  change our minds.}  We emphasize that this consistency is not a
priori obvious because \emph{both $m$ and $d$} must be expanded about
$8$ while maintaining the equality $m=d$. As we shall explicitly
verify, this consistency holds equally well on the levels of the final
$\epsilon$-expansion results, the counter-terms and the Feynman
integrals. This demonstrates that the recent claim by de Albuquerque and
Leite \cite{dAL02}, who asserted that the critical exponents of the
isotropic Lifshitz point cannot be obtained from those for general
$m$, is unfounded.

\section{Two-loop results for the isotropic case $\boldsymbol{m=d}$}
\label{sec:2l}

In the isotropic case $m=d$, only the parallel part of space remains,
so that $\boldsymbol{x}_\|=\boldsymbol{x}$. It is convenient to use
the rescaled variables
\begin{equation}
  \label{eq:varphidef}
  \bphi(\boldsymbol{x})
   =\sigma_0^{-d/8}\,\bpsi(\boldsymbol{y})\;,\quad
   \boldsymbol{y}=\sigma_0^{-1/4}\boldsymbol{x}\;.
\end{equation}
Then the Hamiltonian implied by equation (\ref{eq:Ham}) for $m=d$
becomes
\begin{equation}\label{eq:Hiso}
{\mathcal{H}}^{(\rm iso)}=\int d^dy\,\left[
\frac{1}{2}\,(\triangle\bpsi)^2+{\lambda_0\over 2}\,{({\nabla}
 \bpsi )}^2+
\frac{\tau_0}{2}\,\bpsi^2+\frac{g_0}{4!}\,|\bpsi|^4
\right]\end{equation}
with
\begin{equation}
  \label{eq:r0g0}
  g_0=\sigma_0^{-d/4}\,u_0\;,\quad \lambda_0=\sigma_0^{-1/2}\,\rho_0\;.
\end{equation}
We use dimensional regularization to regularize the ultraviolet (uv)
singularities of the $N$-point cumulants
$G^{(N)}=\langle\psi\ldots\psi\rangle^{\rm cum}$ and vertex
functions $\Gamma^{(N)}$ of the theory in $d=8-\epsilon_8$ dimensions.
These uv singularities can be absorbed via the re-parametrizations
\begin{equation}
  \label{eq:ZpsiZgintro}
\bpsi=Z_\psi^{1/2}\,\bpsi_{\rm ren}\;,
\quad g_0\,F_d=\kappa^{\epsilon_8}\,Z_g\,g\;,
\end{equation}
and
\begin{equation}
  \label{eq:ZtZlambdaintro}
 \tau_0-\tau_{\rm LP}=\kappa^4\,Z_t\,t\;,\quad 
\lambda_0-\lambda_{\rm LP}=\kappa^2Z_\lambda\,\lambda\;,
\end{equation}
where $\tau_{LP}$ and $\lambda_{\rm LP}$, the values of $\tau_0$ and
$\lambda_0$ at the Lifshitz point, vanish in our perturbative approach
based on dimensional regularization. For the normalization constant in
equation (\ref{eq:ZpsiZgintro}) we use the choice
\begin{equation}
F_d={2^{1-d}\,\pi^{-d/2}\,\Gamma(5-\frac{d}{2})\,
\Gamma(\frac{d}{2}-2)^2\over
\Gamma(d-4)}\;.
\end{equation}

In order to determine the renormalization factors $Z_\psi$, $Z_g$,
$Z_t$, and $Z_\lambda$, we compute the Laurent expansions of the
vertex functions $\Gamma^{(2)}$, $\Gamma^{(4)}$,
$\Gamma^{(2)}_{\bpsi^2}$, and $\Gamma^{(2)}_{(\nabla\bpsi)^2}$
at the Lifshitz point $t=\lambda=0$, where the latter two involve
insertions of (one half of) the indicated operators $\bpsi^2$ and
${(\nabla\bpsi)^2}$, respectively. The free propagator with which
we work is given by
\begin{equation}\label{eq:freeprop}
G(y)\equiv\int_{\boldsymbol{q}}\tilde{G}(\boldsymbol{q})\,
\e^{i\boldsymbol{q}\cdot\boldsymbol{y}}=\frac{\pi^{-d/2}}{16}\,
\Gamma{\left[{(d-4)/2}\right]}\,
y^{4-d}\;,\quad \tilde{G}(\boldsymbol{q})=\frac{1}{q^4}\;,
\end{equation}
in position and momentum space, respectively. The values of the
one-loop integral
\begin{equation}\label{eq:I2def}
I_2(q)\equiv\int d^dy\,G(y)^2\,\e^{i \boldsymbol{q}\cdot \boldsymbol{y}}\\
\end{equation}
of $\tilde{\Gamma}^{(4)}(\boldsymbol{q}_1,\ldots,\boldsymbol{q}_4)$
(where $q=|\boldsymbol{q}_1+\boldsymbol{q}_2|$, for example)
and the two-loop integral 
\begin{equation}
  \label{eq:I3def}
I_3(q)\equiv \int d^dy\,G(y)^3\,\e^{i \boldsymbol{q}\cdot \boldsymbol{y}}
\end{equation}
of $\tilde{\Gamma}^{(2)}(\boldsymbol{q})$ can be read off from the
Fourier transform of the generalized function
$|\boldsymbol{y}|^\vartheta$, which is 
\begin{equation}
  \label{eq:Fint}
{\int}\!d^dy\,y^\vartheta\,\e^{-i \boldsymbol{q}\cdot  \boldsymbol{y}}=
2^{\vartheta +d}\,\pi^{d/2} \,
{\Gamma{\left({\vartheta+d\over 2}\right)}\over 
\Gamma{\left(-{\vartheta\over 2}\right)}}\,q^{-\vartheta-d}
\end{equation}
according to equation (2) of reference \cite{GS64spez}. The results
are listed in \ref{sec:2lint} along with the other required two-loop
integrals. We choose the renormalization factors such that the uv
poles are minimally subtracted, obtaining
\begin{equation}
 \label{eq:Zpsi}
Z_\psi=1+\frac{n+2}{3}\,\frac{g^2}{160\,\epsilon_8}+\Or(g^3)\;,
\end{equation}
\begin{eqnarray}
  \label{eq:Zg}
\fl
Z_\psi^2Z_g&=&1+
\frac{n+8}{9}\,\frac{3}{2}\,\frac{g}{\epsilon_8}
+
\left[
\left(\frac{n+8}{9}\,\frac{3}{2\,\epsilon_8}\right)^2+3\,
\frac{5n+22}{27}\,\frac{1}{24\,\epsilon_8}\right]g^2
+\Or(g^3)\;,
\end{eqnarray}
\begin{equation}
  \label{eq:Zt}
\fl
Z_\psi Z_t=1+
\frac{n+2}{6}\,\frac{g}{\epsilon_8}+\frac{n+2}{6}\left[
\frac{n+5}{6}\,\frac{1}{\epsilon_8^2}+\frac{1}{24\,\epsilon_8}
\right]g^2+\Or(g^3),
\end{equation}
and
\begin{equation}
  \label{eq:Zlambda}
Z_\psi\,Z_\lambda=1+\frac{n+2}{3}\,\frac{3\,g^2}{32\,\epsilon_8}+\Or(g^3)\;.
\end{equation}

From these results the beta and exponent functions, defined by
\begin{equation}
  \label{eq:betadef}
  \beta_g(g)\equiv\left.\kappa\partial_\kappa\right|_0 g
\end{equation}
and
\begin{equation}
  \label{eq:expfdef}
  \eta_\wp\equiv \left.\kappa\partial_\kappa\right|_0\ln Z_\wp\;,\quad
   \wp=\psi,t,\lambda\;,
\end{equation}
where $\partial_\kappa|_0$ denotes a derivative at fixed values of the
bare variables $\lambda_0$, $\tau_0$ and $g_0$, follow in a
straightforward fashion. They read
\begin{equation}
  \label{eq:betag}
\beta_g(g)=-\epsilon_8\,g
+\frac{n+8}{6}\,g^2
+\frac{41\,n+202}{1080}\,g^3+\Or(g^4)\;,
\end{equation}
\begin{equation}
  \label{eq:etavarphi}
  \eta_\psi(g)=-\frac{n+2}{3}\,\frac{g^2}{80}+\Or(g^3)\;,
\end{equation}
\begin{equation}
  \label{eq:etat}
  \eta_t(g)=\frac{n+2}{6}\,g-\frac{7\,(n+2)}{720}\,g^2
   +\Or(g^3)\;,
\end{equation}
and
\begin{equation}
  \label{eq:etalambda}
  \eta_\lambda(g)=-\frac{7\,(n+2)\,g^2}{120}
   +\Or(g^3)\;.
\end{equation}
The nontrivial root of $\beta_g(g)$ is
\begin{equation}
  \label{eq:gstar}
g^*=\frac{6\,\epsilon_8}{n+8}-\frac{41\,n+202}{5\,(n+8)^3}\,\epsilon_8^2
+\Or(\epsilon_8^3)\;.
\end{equation}
By evaluating the exponent functions at the fixed point $g^*$, we
arrive at the dimensionality expansions of the exponents
\begin{equation}
  \label{eq:eta}
\eta=\eta_\psi(g^*)=-\frac{3\,(n+2)\,\epsilon_8^2}{20\,(n+8)^2}
+\Or(\epsilon_8^3),
\end{equation}
\begin{eqnarray}
  \label{eq:nu}\fl
\nu=\frac{1}{4+\eta_t(g^*)}
=\frac{1}{4}+\frac{n+2}{n+8}\,{\epsilon_8\over 16}+
\frac{(n+2)(15\,n^2+89\,n+4)\,\epsilon_8^2}{960\,(n+8)^3}
 +\Or(\epsilon_8^3)\;,
\end{eqnarray}
\begin{equation}
  \label{eq:varphi}\fl
  \varphi=\frac{2+\eta_\lambda(g^*)}{4+\eta_t(g^*)}=\frac{1}{2}+\frac{n+2}{n+8}\,\frac{\epsilon_8}{8}
+\frac{(n+2)\,(15\,n^2-163\,n-2012)\,\epsilon_8^2}{480\,(n+8)^3}
+\Or(\epsilon_8^3)\,,
\end{equation}
and
\begin{equation}
  \label{eq:betaq}
  \beta_q= \frac{1}{2}+\frac{21\,(n+2)\,\epsilon_8^2}{40\,(n+8)^2}
+\Or(\epsilon_8^3)\;.
\end{equation}
Likewise we find for the  correction-to-scaling exponent:
\begin{equation}
  \label{eq:omegag}
\omega=\left.\frac{\partial\beta_g(g)}{\partial g}\right|_{g=g^*}
=\epsilon_8 + {\frac{\left(41\,n + 202 \right) \,{\epsilon_8^2}}
    {30\,{{\left( n + 8 \right) }^2}}} + 
  \Or(\epsilon_8^3)\;.
\end{equation}

The results (\ref{eq:eta}) and (\ref{eq:nu}) were already obtained in
reference \cite{HLS75b}\footnote{When referring to the results of the
  work \cite{HLS75b} at the bottom of p.~355 of our previous paper
  \cite{SD01}, we erroneously said that these authors had obtained the
  expansions of the correlation exponents $\nu_{l2}$ and $\eta_{l4}$
  and the correlation-length exponents $\nu_{l2}$ and $\nu_{l4}$ to
  second order in $\epsilon_8$. The alert reader will have noticed
  that the cited computation was concerned merely with the isotropic
  Lifshitz point, so that only $\eta_{l4}=\eta$ and $\nu_{l4}=\nu$
  should have been mentioned.}, but the remaining ones
(\ref{eq:varphi})--(\ref{eq:omegag}) not.

\section{Comparison with the results for the $\boldsymbol{m}$-axial
  Lifshitz point}
\label{sec:comp}
\subsection{Critical, crossover, and correction-to-scaling exponents}
\label{sec:critexp}

In reference \cite{SD01}, which hereafter is referred to as I, we
  considered the critical exponents as functions $f(m,d)$ of $m$ and
  $d$, and determined their expansions in $d$ about $d^*(m)$ at
  \emph{fixed} $m$ to second order in $\epsilon=d^*(m)-d$. For
  $d=m=8-\epsilon_8$, we have
\begin{equation}
  \label{eq:eps8}
  \epsilon={\epsilon_8/ 2}\;.
\end{equation} 
The series expansion coefficients of the terms of zeroth and first order in
$\epsilon$, found in I, are independent of $m$. To check the consistency of the
above results with those of references \cite{SD01,DS01b}, we therefore
must merely substitute the limiting values of the latters'
$\Or(\epsilon^2)$ terms for $m\to 8-$, and use the relation (\ref{eq:eps8}).

The exponents $\eta$, $\nu$, and $\omega$ given in
equations (\ref{eq:eta}), (\ref{eq:nu}), and (\ref{eq:omegag}) read in
the notation of I
\begin{equation}
  \label{eq:expcorr}
  \eta=\eta_{l4}(d,d)\;,\quad \nu=\nu_{l4}(d,d)\;,\quad
\omega=\omega_{l4}(d,d)\;,
\end{equation}
respectively. The $\epsilon$ expansions of $\eta_{l4}$ and $\nu_{l4}$
are given in equations (I.62) and (I.64), and the one for
$\omega_{l4}$ follows from equations (I.66)--(I.68), where (I.xx)
denotes the equation (xx) of I. Note that the contribution $\propto
\lim_{m\to 8-}j_\phi(m)/(8-m)=-j_\phi'(8-)$ which is present in the
$\epsilon$ expansion (I.66) of the anisotropy exponent $\theta$, does
not contribute to the ratio $\omega_{l4}=\omega_{l2}/\theta$ at
$\Or(\epsilon^2)$. The only term involving $j_\phi(m)$ at
$\Or(\epsilon^2)$ is directly proportional to $j_\phi(m)$. Since the
latter integral is of order $\epsilon_8$, all contributions to the
series expansion of the correction-to-scaling exponent $\omega$ to
$\Or(\epsilon_8^2)$ that originate from $j_\phi(m)$ vanish, as they
should.

Upon substituting the values (I.86) of the integrals $j_\phi(8-)$,
\ldots, $J_u(8-)$ into the $\Or(\epsilon^2)$ coefficients, the
resulting series expansions to order $\epsilon^2_8$ of the exponents
(\ref{eq:expcorr}) reduce to the above results (\ref{eq:eta}),
(\ref{eq:nu}) and (\ref{eq:omegag}). Likewise, the $\epsilon_8$
expansions of $\varphi=(1+\eta_\rho^*)/(2+\eta_\tau^*)$ and
$\beta_q=\nu_{l4}/\varphi$ that are implied by equations (I.63) ---
(I.65) for the fixed-point quantities $\eta_\tau^*$, $\eta_\sigma^*$,
and $\eta_\rho^*$ of I agree with equations (\ref{eq:varphi}) and
(\ref{eq:betaq}). The remaining critical exponents, such as the
specific-heat exponent $\alpha$, the order-parameter exponent $\beta$,
and the susceptibility exponent $\gamma$, are related to $\eta$ and
$\nu$ via known scaling and hyperscaling relations. Thus, there is no
need to explicitly verify their consistency with the results of I.

For the case of an $m$-axial Lifshitz point a variety of other
physically meaningful critical exponents can be defined, such as the
exponent $\eta_{l2}$ (which governs the momentum dependence of the
inverse correlation function
$\tilde{\Gamma}^{(2)}({\boldsymbol{q}}_\|{=}
\boldsymbol{0},\boldsymbol{q}_\perp)\sim q_\perp^{2-\eta_{l2}}$ at the
Lifshitz point), the parallel correlation-length exponent $\nu_{l2}$,
the related anisotropy exponent $\theta=\nu_{l4}/\nu_{l2}$, and the
correction-to-scaling exponent $\omega_{l2}=\theta\,\omega_{l4}$.  The
$\epsilon$ expansions to order $\epsilon^2$ of all these exponents
have been given in I. These latter exponents are not needed in the
case of the isotropic Lifshitz point. The situation is complementary
to the $m=0$ case of the standard $|\bphi|^4$ theory, where exponents
requiring the parallel part of space (such as $\eta_{l4}$ and
$\nu_{l4}$) loose their physical significance. Remarks analogous to
those made in I apply here: The limits $d-m\to 0$ and $m\to 0$ of
those exponents that are not required in the isotropic cases $m=d$ and
$m=0$ may well exist, and it is conceivable that their limiting values
will turn out to have significance for certain problems of statistical
physics. We will not pursue this question further here. Note, however,
that the limiting value of $\eta_{l2}$ for $m\to d$ appears to exist
and to be finite since its $\epsilon^2$ term involves the integral $j_\phi(m)$ in the combination $j_\phi(m)/(8-m)$,  according to equation
(I.61). Since $j_\phi(8-)=0$, the  limit $m\to 8-$  of this
combination is nonsingular and nonzero provided that $j_\phi'(8-)\ne 0$ (as
we expect). For the same reason, the contribution $\propto
j_\phi(m)/(8-m)$ to $\nu_{l2}$ in equation (I.63) remains finite in
this limit, as do the other contributions to the $\Or(\epsilon^2)$
term.

\subsection{Comparison of renormalization functions}
\label{sec:renfac}

In the foregoing subsection we have shown that the correct series
expansions in powers of $\epsilon_8$ of the critical, crossover, and
correction-to-scaling exponents of the isotropic Lifshitz point follow
from the results of I. An analogous result holds for other universal
quantities. This is because the required renormalization factors
$Z_\psi$, $Z_g$, $Z_t$, and $Z_\lambda$, and hence the implied beta
and exponent functions $\beta_g(g)$ and $\eta_\psi$, $\eta_t$,
$\eta_\lambda$ may all be obtained from the $m$-dependent results of
I.

A direct way of seeing this is to compare the Feynman integrals on
which the analysis of the present paper is based (and which are
computed in \ref{sec:2lint}) with their analogues of I.  To facilitate
this comparison, it is advisable to get rid of the choices of the
factors $F_d$ and $F_{m,\epsilon}$ we absorbed in $g$ and $u$
(the renormalized coupling constant of I), respectively.
To achieve this goal, we must merely divide two-loop
Feynman integrals such as $I_3$, $I_4$, and $I_5$ by the square of
$\epsilon_8\,I_2$ [as can be seen from equations (\ref{eq:I2res}) and
(I.38)]. That is, the normalized integral
$I_4(\boldsymbol{e},\boldsymbol{0})/[\epsilon_8\,
I_2(\boldsymbol{e})]^2$ should be compared with the quantity
$\{I_4(\boldsymbol{e}_\perp,\boldsymbol{0})/[\epsilon\,
I_2(\boldsymbol{e}_\perp)]^2\}_{m=8-\epsilon_8,\epsilon=\epsilon_8/2}$,
where $\boldsymbol{e}$ and $\boldsymbol{e}_\perp$ are unit $d$ and
$d-m$ vectors, respectively. Utilizing equations (I.B.14), (I.86) and
(I.B.13) of I, and the corresponding results (\ref{eq:I2res}) and
(\ref{eq:I4exp}) for the isotropic case, one finds that the pole parts
of these quantities agree indeed:
\begin{equation}
\label{eq:I4comp}\fl
 \left. {I_4(\boldsymbol{e}_\perp,\boldsymbol{0})\over
     [I_2(\boldsymbol{e}_\perp)]^2}\right|_{m=8-e8,\epsilon=\epsilon_8/2}=
\frac{1}{2\,\epsilon_8}\,{\left[
\frac{1}{\epsilon_8}-\frac{1}{12}+\Or(\epsilon_8)
\right]}={I_4(\boldsymbol{e},\boldsymbol{0})\over [\epsilon\,
I_2(\boldsymbol{e})]^2}\;.
\end{equation}
Analogous results hold for the other integrals needed to determine the
two-loop counter-terms. We have
\begin{eqnarray}\fl
  \label{eq:I3comp}
 {I_3(\boldsymbol{q})\over [\epsilon_8\,
I_2(\boldsymbol{e})]^2}=\frac{3\,q^4}{80\,\epsilon_8}+\Or(\epsilon_8^0)\nonumber\\
\lo =
\left\{\left[\frac{F_{m,\epsilon}^2}{\epsilon}\,\frac{j_\sigma(m)\,q^4}{16\,m(m+2)}
+\Or(\epsilon^0)\right] \frac{1}{[\epsilon\,
I_2(\boldsymbol{e}_\perp)]^2}\right\}_{m=8-\epsilon_8,\epsilon=\epsilon_8/2}
\end{eqnarray}
and
\begin{eqnarray}\fl
  \label{eq:I5comp}
 {I_5(\boldsymbol{q})\over [\epsilon_8\,
I_2(\boldsymbol{e})]^2}=\frac{-3\,q^2}{16\,\epsilon_8}+\Or(\epsilon_8^0)\nonumber\\
\lo =
\left\{\left[\frac{-F_{m,\epsilon}^2}{\epsilon}\,\frac{j_\rho(m)}{4\,m}\,q^2
+\Or(\epsilon^0)\right] \frac{1}{[\epsilon\,
I_2(\boldsymbol{e}_\perp)]^2}\right\}_{m=8-\epsilon_8,\epsilon=\epsilon_8/2}\;,
\end{eqnarray}
where $F_{m,\epsilon}=\epsilon\,I_2(\boldsymbol{e}_\perp)$ is the
counterpart of the normalization constant $F_d$. The analogues of
$I_3$ and $I_5$ we inserted on the right-hand sides of these equations
can be read off from equations (I.B.4) and (I.B.6). Thus the pole
terms found in I reduce in the isotropic case precisely to those
obtained here. Provided we use the same conventions as here to fix the
counter-terms, we get identical results for the $Z$ factors $Z_\psi$,
$Z_g$, $Z_t$, and $Z_\lambda$.

Let us briefly comment on the fact that we evaluated the integrals
$I_2$ and $I_4$ in I at a momentum $\boldsymbol{q}$ with vanishing parallel
component $\boldsymbol{q}_\|$ and finite perpendicular one
$\boldsymbol{q}_\perp$. In the limit $m\to d$, the latter becomes
zero-dimensional. However, this is no cause of concern. First of all,
we should get reasonable results in this limit (as our results in I
show) because the opposite would indicate a failure of the analytic
continuation of the momentum-space integrals $\int d^{d-m}q$ in $d-m$.
Second, taking $\boldsymbol{q}=(\boldsymbol{0},\boldsymbol{q}_\perp)$
may be viewed as the special choice
$\exp(i\boldsymbol{q}_\perp\cdot\boldsymbol{x}_\perp -a\, x^2)$, with
$a\to 0+$, of a test function utilized to determine the pole term of
the corresponding distributions $G(\boldsymbol{x})^2$ and the Fourier
back-transform of $I_4$ to position space. The result
$G(\boldsymbol{x})^2\,F_{m,\epsilon}^{-2}=
\epsilon^{-1}\,\delta(\boldsymbol{x})+\Or(\epsilon^0)$ we obtained in
I can equally well be derived by computing the action of $G^2$ on
other test functions (integrating over a finite sphere centered at the
origin would do). If we set $d=m=8-\epsilon_8$ in this result, we
recover the correct pole term of the isotropic case. Needless to say
that only the ($m=d$)-dimensional part $\delta(\boldsymbol{x}_\|)$ of
the delta function remains for $d=m$, where
$\boldsymbol{x}=\boldsymbol{x}_\|$.

By comparing the renormalized action implied by our
re-parameterizations (\ref{eq:ZpsiZgintro}) and
(\ref{eq:ZtZlambdaintro}) with the renormalized action of I when
$m=d$, one can easily see how our two-loop results
(\ref{eq:Zpsi})--(\ref{eq:Zlambda}) for these renormalization factors
can be expressed in terms of those of I. Since to order $u^2$ only the
terms $\propto u^2/\epsilon$ depend on $m$, we can set $m=8$ when
evaluating them. The desired relations then become
\begin{equation}
\label{eq:Zpsict}
  Z_\psi(g,\epsilon_8)=\left[
Z_\sigma\,Z_\phi
\right]_{u=g/2;m=8,\epsilon=\epsilon_8/2}+\Or(g^3)\;,
\end{equation}
\begin{equation}
\label{eq:Zgct}
  {\left(Z_\psi^2\,Z_g\right)}(g,\epsilon_8)=
\left[Z_u\,Z_\phi^2\,Z_\sigma^{m/4}
\right]_{u=g/2;m=8,\epsilon=\epsilon_8/2}+\Or(g^3)\;,
\end{equation}
\begin{equation}
\label{eq:Ztcr}
  {\left(Z_\psi\,Z_t\right)}(g,\epsilon_8)=
\left[Z_\phi\,Z_\tau
\right]_{u=g/2;m=8,\epsilon=\epsilon_8/2}+\Or(g^3)\;,
\end{equation}
and
\begin{equation}
\label{eq:Zlambdact}
  {\left(Z_\psi\,Z_\lambda\right)}(g,\epsilon_8)=
\left[Z_\rho\,Z_\phi\,Z_\sigma^{1/2}
\right]_{u=g/2;m=8,\epsilon=\epsilon_8/2}+\Or(g^3)\;,
\end{equation}
where the replacement $u\to g/2$ is due to the different normalization
constants whose ratio is
$F_{m=8-\epsilon_8,\epsilon_8/2}/F_d={1/2}+\Or(\epsilon_8)$.

For the associated RG functions, this translates into
\begin{equation}
  \label{eq:betagu}
  \beta_g(g;\epsilon_8)
   =-\epsilon_8\,g+4\,\beta_u(g/2;m=8,\epsilon=0)+\Or(g^3)\,,
\end{equation}
\begin{equation}
  \label{eq:etagu}\fl
  \eta_\psi(g)=2\,\eta_\sigma(g/2;m=8)+\Or(g^3)\;,\quad
  \eta_t(g)=2\,\eta_\tau(g/2;8)+\Or(g^3)\;,
\end{equation}
and
\begin{equation}
  \label{eq:etalambdagu}
  \eta_\lambda(g) =2\,\eta_\rho(g/2;8)-\eta_\sigma(g/2;8)+\Or(g^3)\;.
\end{equation}
The reader is invited to convince himself that the results
(\ref{eq:Zpsi})--(\ref{eq:Zlambda}) for the $Z$ factors and the RG
functions (\ref{eq:betag})--(\ref{eq:etalambda}) of the isotropic
Lifshitz point are indeed recovered from the above formulae when the
$m$-dependent results of I are inserted into them, utilizing the
required values (I.86) of the integrals $j_\phi(m)$, \ldots, $J_u(m)$
at $m=8$.

\section{Concluding remarks}
\label{sec:concl}

In summary, we have shown that the counter-terms, RG functions, and the
series expansions in powers of $\epsilon_8-8-d$ of the critical,
crossover, and correction-to-scaling exponents of the isotropic
Lifshitz point can be obtained from the results of I for the more
general case of an $m$-axial Lifshitz point. We have verified their
validity by performing an independent two-loop calculation directly
for the isotropic case $d-m=0$, using well-established standard
field-theory techniques.

The fact that a proper field-theoretic analysis of the critical
behaviour at $m$-axial Lifshitz points, based on the $\epsilon$
expansion, covers also the isotropic case $m=d$, should not be too
surprising: After all, the free propagator $G(\boldsymbol{x})$ for
general $m\ne d$ goes smoothly over into the correct isotropic one of
equation (\ref{eq:freeprop}) as $m-d \to 0$. The Feynman graphs of the
primitively divergent vertex functions are distributions constructed
from $G(\boldsymbol{x})$ having well-defined Laurent expansions in
$\epsilon$. The renormalization factors are determined by the pole
parts of the final subtractions which the primitively divergent vertex
functions require. Clearly, their behaviour in the isotropic limit
should comply with that of $G(\boldsymbol{x})$.

In closing, let us straighten out another critique by de Abuquerque
and Leite \cite{dAL02}. They claimed that our two-loop calculation was
incomplete because we set the parallel component $\boldsymbol{Q}_\|$
of the momentum $\boldsymbol{Q}$ to zero when computing the two-loop
integral $I_4(\boldsymbol{Q},\boldsymbol{K})$ of the graph
$\raisebox{-8pt}{\begin{texdraw}
\drawdim pt \setunitscale 2.0   \linewd 0.3
\move(-7 -1.5)\rlvec(15 6)
\move(-7 1.5)\rlvec(15 -6)
\move(5 0)
\lellip rx:1 ry:3 \move(8 0)
\end{texdraw}}$.
However, the \emph{final} subtractions we determined in this part of our
computation were those associated with the $|\bphi|^4$ counter-term
and the vertex function $\Gamma^{(2)}_{\bphi^2}$, both of which are
\emph{momentum-independent}. It is true that
$I_4(\boldsymbol{Q},\boldsymbol{K})$ has first order poles in
$\epsilon$ that depend also on $\boldsymbol{Q}_\|$. These are induced by
the divergent sub-integral
$\raisebox{-2.2pt}{\begin{texdraw}
\drawdim pt \setunitscale 2.0   \linewd 0.3
\lellip rx:4 ry:1.8
\move(4 0)\rlvec(2 2)
\move(4 0)\rlvec(2 -2)
\move(-4 0)\rlvec(-2 -2)
\move(-4 0)\rlvec(-2 2)\move(-7.5 0) \move(7.5 0)
\end{texdraw}}$,
and are cancelled automatically by the subtractions provided by the
one-loop counter-terms, in renormalizable theories such as the ones
considered in I and here. Thus the convenient choice
$\boldsymbol{Q}_\|=0$ is possible. Readers who would like to see
this re-iterated and verified explicitly in the present context may
consult \ref{sec:canc}.

\ackn Our work on which this paper is based was supported via the
Leibniz programme DI 378/2-1 of the Deutsche Forschungsgemeinschaft.

\appendix

\section{Feynman integrals}
\label{sec:2lint}

Equation (\ref{eq:Fint}) yields for the integrals (\ref{eq:I2def}) and
(\ref{eq:I3def}) the results
\begin{equation}
  \label{eq:I2res}
I_2(q)
={2^{-d}}\,{{\pi }^
       {-{\frac{d}{2}}} }\,{\frac{{q^{d-8}}\,
     {\Gamma}(4 - 
       {\frac{d}{2}})\,
     {{{\Gamma}(
          {\frac{d}{2}}-2)}^2}}{
     {\Gamma}(d-4)}}=F_d\,\frac{q^{-\epsilon_8}}{\epsilon_8}
\end{equation}
and
\begin{eqnarray}
  \label{eq:I3res}
I_3(q)
&=&F_d^2\,
{\frac{{q^
       {2\,\left(d-6 \right) }}\,{\Gamma}(6 - d)\,
     {{{\Gamma}( d-4)}^2}
     }{4\,{{{\Gamma}(
         5 - {\frac{d}{2}})}^2}\,
     {\Gamma}({\frac{d}{2}}-2)\,
     {\Gamma}( {\frac{3\,d}{2}}-6)}}\\
&=&F_d^2\,q^4\,{\left[
\frac{3}
    {80\,\epsilon_8} + 
  {\frac{211 - 
        240\,\ln q }{
      3200}} + \Or(\epsilon_8)\right]}\;,
\end{eqnarray}
respectively. The two-loop graph
\raisebox{-8pt}{\begin{texdraw}
\drawdim pt \setunitscale 2.0   \linewd 0.3
\move(-7 -1.5)\rlvec(15 6)
\move(-7 1.5)\rlvec(15 -6)
\move(5 0)
\lellip rx:1 ry:3
\end{texdraw}}
involves the Feynman integral
\begin{eqnarray}
I_4(\boldsymbol{q}_{12},\boldsymbol{q}_3)&\equiv&
\int_{\boldsymbol{q}}\, 
\frac{1}{q^4\,|\boldsymbol{q}-\boldsymbol{q}_{12}|^4}\,
I_2(|\boldsymbol{q}+\boldsymbol{q}_3|)\;.
\end{eqnarray}
Using a familiar trick due to Feynman we can transform the product of
denominators into a single power by the introduction of two Feynman
parameters $s$ and $t$. Performing the integral over $\boldsymbol{q}$
then gives
\begin{eqnarray}
\label{eq:I4res}
\fl 
I_4(\boldsymbol{q}_{12},\boldsymbol{q}_3)=
{F_d\over \epsilon_8}\,
\frac{\Gamma(\epsilon_8)}{(4\pi)^{d/2}\,\Gamma(\frac{\epsilon_8}{2})}
\,\int_0^1ds\,s(1-s)\,
\int_0^1dt\,t^3\,(1-t)^{-1+\epsilon_8/2}\nonumber\\ \lo \times
\,\Big\{s\,t\,q_{12}^2
+(1-t)\,q_3^2-[(1-t)q_3-s\,t\,q_{12}]^2\Big\}^{-\epsilon_8}\;.
\end{eqnarray}
By adding and subtraction the value of the integrand at $q_3=0$, one
easily sees that the pole part of
$I_4(\boldsymbol{q}_{12},\boldsymbol{q}_3)$ is independent of
$\boldsymbol{q}_3$ (noting that the pre-factor of the integral on the
right-hand side is of order $\epsilon_8^{-1}$). We therefore set
$q_3=0$, which leads us to the integral
\begin{equation}
  \label{eq:Jd}
\fl
J(d) \equiv
\int_0^1ds\,s^{1-\epsilon_8}\,(1-s)\,
\int_0^1dt\,t^{3-\epsilon_8}\,(1-t)^{-1+{\epsilon_8}/{2}}
\,(1-s\,t)^{-\epsilon_8}\;,
\end{equation}
whose dimensionality expansion
\begin{equation}
  \label{eq:Jdexp}
J(8-\epsilon_8) =\frac{1}{3\,\epsilon_8}+\frac{1}{4}+\Or(\epsilon_8)
\end{equation}
can be determined in a straightforward fashion. Upon expanding the
pre-factor of the integral, one arrives at the result
\begin{equation}\label{eq:I4exp}
I_4(\boldsymbol{q}_{12},\boldsymbol{q}_3)=
F_d^2\,{q_{12}^{-2\,\epsilon_8}\over 2\,\epsilon_8}\,
{\left[\frac{1}{\epsilon_8}-\frac{1}{12}+\Or(\epsilon_8)
\right]}\;.
\end{equation}

As usual, the required two-loop integrals of $\Gamma^{(2)}_{\bphi^2}$
involve just the integrals $I_2$ and $I_4$. However, we 
also need the integral associated with the two-loop graph
\raisebox{-4.25pt}{
\begin{texdraw}
\drawdim pt \setunitscale 2.0   \linewd 0.3
\lellip rx:5 ry:2.8
\move(-8 0)
\lvec(8 0)
\move(0 2.8)
\fcir f:0 r:0.7
\rmove(-1.15 -1)\rlvec(0 2)
\rmove(2.25 -2)\rlvec(0 2) \move(9.5 0)
\end{texdraw}}
of
$\tilde{\Gamma}^{(2)}_{(\nabla\bpsi)^2}(\boldsymbol{q},\boldsymbol{Q})$,
where $\boldsymbol{Q}$ denotes the momentum of the inserted operator
$[(\nabla\bpsi)^2/2]_{\boldsymbol{Q}}$. Since the limit
$\boldsymbol{Q}\to \boldsymbol{0}$ of this Feynman integral is not
infrared-singular and exists, we can consider directly the case
$\boldsymbol{Q}=\boldsymbol{0}$. The upper line with such an insertion
corresponds in position space to the function
\begin{equation}
 D(y)= \int_{\boldsymbol{q}}
       \frac{\e^{i\boldsymbol{q}\cdot\boldsymbol{y}}}{q^6}
     =\frac{\pi^{-d/2}}{128}\,\Gamma{\left(\frac{d-6}{2}\right)}\,y^{6-d}\;.
\end{equation}
Thus the required Feyman integral can be written as
\begin{equation}
  \label{eq:I3prime}
  I_5(q)={\int}\!d^dy\,G(y)^2\,D(y)\,
          \e^{i\boldsymbol{q}\cdot\boldsymbol{y}}\;. 
\end{equation}
It can be evaluated in a straightforward fashion to obtain
\begin{eqnarray}
  \label{eq:I5}
  I_5(q)/F_d^2&=&\frac{\Gamma(7-d)\,\Gamma[(d-6)/2]\,
\Gamma(d-4)^2}{8\,\Gamma[(10-d)/2]^2\,\Gamma[(d-4)/2]^2\,
\Gamma[(3d/2)-7]}\,q^{2(d-7)}\nonumber\\
&=&-\frac{3\,q^2}{16\,\epsilon_8}+\Or(\epsilon_8^0)\;.
\end{eqnarray}

Utilizing the Laurent expansions of the integrals given above, one can
derive the results (\ref{eq:Zpsi})--(\ref{eq:Zlambda}) for the
renormalization factors in a straightforward manner.

\section{Cancellation of momentum-dependent pole terms}
\label{sec:canc}

In this appendix we show explicitly for the case of the $m$-axial
Lifshitz point that the momentum-dependent poles of the two-loop graph
of $\Gamma^{(4)}$ appearing first on the the right-hand side of
equation (\ref{eq:rq}) are cancelled by the subtractions which the
one-loop counter-terms provide for the divergent sub-integral inside the
dashed box. It is sufficient to consider the problem for a
one-component order parameter ($n=1$). The lines correspond to the
free propagator of the Hamiltonian (\ref{eq:Ham}) for
$\tau_0=\rho_0=0$; its explicit forms in momentum and position space
can be found in equations (I.5)--(I.11).

Zimmermann's forest formula \cite{Zim70} tells us, for any Feynman
graph, into which subtractions for its divergent subgraphs the
lower-order counter-terms translate. In our case, the result is
particularly simple: The one-loop $|\bphi|^4$ counter-term yields the
subtraction
\begin{equation}
  \label{eq:rq}
  \overline{\mathcal R}{\left[\,
\raisebox{-8pt}{\begin{texdraw}
\drawdim pt \setunitscale 2.5   \linewd 0.2
\move(-7 -1.5)\rlvec(15 6)
\move(-7 1.5)\rlvec(15 -6)
\move(5 0)
\lellip rx:1 ry:3 \move(8 0)
\end{texdraw}}\,\right]}= 
\raisebox{-8pt}{\;\begin{texdraw}
\drawdim pt \setunitscale 2.5   \linewd 0.2
\move(-7 -1.5)\rlvec(15 6)
\move(-7 1.5)\rlvec(15 -6)
\move(5 0)
\lellip rx:1 ry:3 \move(8 0)
\end{texdraw}\;}
-
\;\raisebox{-8pt}{\begin{texdraw}
\drawdim pt \setunitscale 2.5   \linewd 0.2
\move(-7 -1.5)\rlvec(15 6)
\move(-7 1.5)\rlvec(15 -6)
\move(5 0)
\lellip rx:1 ry:3 \move(8 0) 
\lpatt(1 1)\linewd 0.2  \move(3 -4) 
\rlvec(0 8) \rlvec(4 0) \rlvec(0 -8) \rlvec(-4 0)
\end{texdraw}}\;.
\end{equation}
Here the notation $\overline{\mathcal R}[\gamma]$ is used to indicate
the quantity that results from a Feynman integral $I[\gamma]$ of a
graph $\gamma$ by making the required subtractions for all its
divergent subgraphs but not the final subtraction for $\gamma$ itself.
The boxed subgraph denotes its singular part.  This is local in
position space (namely, proportional to $\epsilon^{-1}$ times the
$|\bphi|^4$ vertex). Thus the subtracted graph involves (besides a
momentum-independent second-order pole) a momentum-dependent
first-order pole of the form $\epsilon^{-1}$ times the
momentum-dependent term of order $\epsilon^0$ of the graph that
results when the box is contracted to a point. The latter is again the
one-loop graph of $\Gamma^{(4)}$.

The pole terms of the difference (\ref{eq:rq}) must be
momentum-independent and local in position space so that they can be
absorbed via the $|\bphi|^4$ counter-term. From equation (I.38) we
know that the singular part of the divergent sub-integral $I_2$ of
$I_4$ agrees with the pole part of
$I_2(\boldsymbol{e}_\perp)=F_{m,\epsilon}\,\epsilon^{-1}$, where
$F_{m,\epsilon}$ is the constant (I.39) while $\boldsymbol{e}_\perp$
denotes a unit $d-m$ vector. Hence the difference of Feynman integrals
we are concerned with becomes
\begin{equation}
  \label{eq:rI4}\fl
 \overline{\mathcal R}[I_4](\boldsymbol{Q},\boldsymbol{K})=
\int_{\boldsymbol{q}}\frac{1}{q_\perp^2+q_\parallel^4}\,
\frac{1}{|\boldsymbol{q}_\perp+\boldsymbol{Q}_\perp|^2
+|\boldsymbol{q}_\|+\boldsymbol{Q}_\||^4}\,
{\left[I_2(\boldsymbol{q}-\boldsymbol{K})-
I_2(\boldsymbol{e}_\perp)
\right]}\;.
\end{equation}
We must show that the difference
$\overline{\mathcal R}[I_4](\boldsymbol{Q},\boldsymbol{K})
-\left.\overline{\mathcal R}[I_4](\boldsymbol{Q},\boldsymbol{K})
\right|_{\boldsymbol{Q}_\|=\boldsymbol{0}}$ is regular in
$\epsilon$. This difference is given by the analogue of equation
(\ref{eq:rI4}) that results through the replacement
\begin{equation}\fl
  \label{eq:intrep}
  \frac{1}{{\big|\boldsymbol{q}_\perp+\boldsymbol{Q}_\perp\big|}^2
+{\big|\boldsymbol{q}_\|+\boldsymbol{Q}_\|\big|}^4}\to 
 \frac{-Q_\|^2\,(Q_\|^2+2q_\|^2)-4\,\big(\boldsymbol{Q}_\|\cdot\boldsymbol{q}_\|+q_\|^2+Q_\|^2\big)\,\boldsymbol{Q}_\|\cdot\boldsymbol{q}_\|}{{\big[|\boldsymbol{q}_\perp+\boldsymbol{Q}_\perp|^2
+|\boldsymbol{q}_\|+\boldsymbol{Q}_\||^4\big]}{\big[
|\boldsymbol{q}_\perp+\boldsymbol{Q}_\perp|^2
+q_\|^4\big]}}\;.
\end{equation}
The subtraction has improved the uv behaviour of the integral by two
powers of $q_\|$, making it uv convergent at the upper critical
dimension in accordance with
Weinberg's theorem \cite{Wei60}. Hence the momentum-dependent
poles---and especially the $\boldsymbol{Q}_\|$-dependent
ones---cancel, as they should, both in $\Gamma^{(4)}$ as well as in
$\Gamma^{(2)}_{\bphi^2}$.

\section*{References}

\end{document}